\documentclass[journal]{IEEEtran}
\usepackage{cite}

\ifCLASSINFOpdf
   \usepackage[pdftex]{graphicx}
\else
\fi
\usepackage{amsmath}
\usepackage[cmintegrals]{newtxmath}
\usepackage{bm}
\usepackage{algorithm}
\usepackage{algpseudocode}

\usepackage{array}
\usepackage{booktabs}
\usepackage{multirow,multicol}

\ifCLASSOPTIONcompsoc
 \usepackage[caption=false,font=normalsize,labelfont=sf,textfont=sf]{subfig}
\else
 \usepackage[caption=false,font=footnotesize]{subfig}
\fi
\usepackage[export]{adjustbox}

\usepackage{url}
\usepackage{hyperref}

\newcommand{\sect}[1]{Section~\ref{sec:#1}}

\newcommand{\eqn}[1]{Eqn.~(\ref{eqn:#1})}
\newcommand{\fig}[1]{Fig.~\ref{fig:#1}}
\newcommand{\tbl}[1]{Table~\ref{tab:#1}}
\newcommand{\alg}[1]{Algorithm~\ref{alg:#1}}

\newcommand{\twosect}[2]{Sections~\ref{sec:#1} and \ref{sec:#2}}

\DeclareMathOperator*{\argmax}{arg\,max}

\begin{document}
%
\title{Combining Frame-Synchronous and Label-Synchronous Systems for Speech Recognition}
%
%
%

\author{Qiujia Li,~\IEEEmembership{Student Member,~IEEE,}
        Chao Zhang,~\IEEEmembership{Member,~IEEE,}
        and~Philip C. Woodland,~\IEEEmembership{Fellow,~IEEE}
\thanks{Q. Li, C. Zhang and P.C. Woodland are with the Department
of Engineering, University of Cambridge, Cambridge,
CB2 1PZ UK e-mail: ( ql264@cam.ac.uk; cz277@cam.ac.uk; pcw@eng.cam.ac.uk).}
\thanks{Manuscript received Month Date, 2021; revised Month Date, 2021.}}

%
%

\markboth{Submitted to IEEE/ACM Transactions on Audio Speech and Language Processing}%
{Li \MakeLowercase{\textit{et al.}}: Combining Frame-Synchronous and Label-Synchronous Systems for Speech Recognition}
%



\maketitle

\begin{abstract}
Commonly used automatic speech recognition (ASR) systems can be classified into frame-synchronous and label-synchronous categories, based on whether the speech is decoded on a per-frame or per-label basis. Frame-synchronous systems, such as traditional hidden Markov model systems, can easily incorporate existing knowledge and can support streaming ASR applications. Label-synchronous systems, based on attention-based encoder-decoder models, can jointly learn the acoustic and language information with a single model, which can be regarded as audio-grounded language models. In this paper, we propose rescoring the $N$-best hypotheses or lattices produced by a first-pass frame-synchronous system with a label-synchronous system in a second-pass. By exploiting the complementary modelling of the different approaches, the combined two-pass systems achieve competitive performance without using any extra speech or text data on two standard ASR tasks. For the 80-hour AMI IHM dataset, the combined system has a 13.7\% word error rate (WER) on the evaluation set, which is up to a 29\% relative WER reduction over the individual systems. For the 300-hour Switchboard dataset, the WERs of the combined system are 5.7\% and 12.1\% on Switchboard and CallHome subsets of Hub5'00, and 13.2\% and 7.6\% on Switchboard Cellular and Fisher subsets of RT03, up to a 33\% relative reduction in WER over the individual systems.
\end{abstract}


%
\IEEEpeerreviewmaketitle

\section{Introduction}
%
%
%
%

\label{sec:intro}

\IEEEPARstart{A}{utomatic} speech recognition (ASR) is a task that converts a speech sequence to its corresponding text sequence. Since the number of frames in the input sequence is generally different from the number of labels in the output sequence, the decoding procedure that searches for the most probable label sequence given the speech sequence can be performed either on a per-frame or a per-label basis. Based on this criterion, most widely used ASR systems can be divided into \textit{frame-synchronous} and \textit{label-synchronous} categories.

Regarding frame-synchronous systems, an extensively studied and widely adopted ASR approach is the noisy source-channel model~\cite{jelinek1997statistical}. It generally consists of an acoustic model (AM) based on hidden Markov models (HMMs), a language model (LM) and a decoder that searches for the most likely word sequence by incorporating acoustic, phonetic, and lexical information. One of the key characteristics of the noisy source-channel model is that the decoding step is synchronous with the input acoustic frames. The emission probabilities of the HMM-based AMs can be modelled by artificial neural networks (NNs)~\cite{bourlard1994connectionist}.

Besides the modular HMM-based systems, there is a long history in the investigation of integrated NN-based systems. A popular approach is called connectionist temporal classification (CTC)~\cite{Graves2006CTC}. CTC can be viewed as a variant of the noisy-source channel model and often uses simpler training and decoding procedures than the NN-HMM systems. 
 
By allowing the model to produce blank symbols and collapsing repetitive symbols, CTC outputs frame-asynchronous label sequences through a frame-synchronous decoding procedure. More recently, the recurrent neural network transducer (RNN-T)~\cite{Graves2012SequenceTW} was proposed to improve CTC by removing the independence assumption in output labels by incorporating an additional prediction network in the model structure. 

Inspired by recent advances in machine translation~\cite{Bahdanau2015NeuralMT}, the attention-based encoder-decoder (AED) models have emerged as another integrated NN-based approach for ASR~\cite{Chorowski2015AttentionBasedMF,Lu2015ASO,Chan2016ListenAA}. Since the attention mechanism captures the alignment between the input frames and output labels, the decoder in AED can operate at a per-label basis, which predicts the current label by attending to multiple input frames and taking the previous labels into account, and is, therefore, a label-synchronous method. There are also various other label-synchronous methods studied in the speech community~\cite{Glass2003CSL,Lee2007AnOO,Zweig2009ASC} and algorithms that allow frame-synchronous systems to operate in a label-synchronous fashion~\cite{Jiang2021VFR}.

Due to their different modelling approaches, frame-synchronous and label-synchronous systems are highly complementary \cite{Prabhavalkar2017ACO,Li2019IntegratingSA,Dong2020ACO}. On one hand, frame-synchronous systems, such as HMM-based systems, more naturally handle streaming data and incorporate structured knowledge such as lexicons. Label synchronous systems, on the other hand, are designed to model the acoustic and language information jointly and can serve as an audio-grounded LM. 
In this paper, we propose combining these two distinctive types of systems in a two-pass fashion to leverage the advantages of each. 
Overall, The contributions of this paper are as follows:
\begin{enumerate}
    \item proposes a simple and effective combination framework for frame-synchronous and label-synchronous systems using $N$-best and lattice rescoring;
    \item proposes an improved lattice rescoring algorithm for label-synchronous systems;
    \item achieves competitive word error rates (WERs) on AMI and Switchboard datasets without additional acoustic and text data.
\end{enumerate}

In the rest of the paper, related work is first described in \sect{related}. Then \sect{fsync} details frame-synchronous systems and draws connections between HMM-based and CTC acoustic models. \sect{lsync} introduces label-synchronous systems and compares between RNN decoders and Transformer decoders. Motivated by the complementarity of two types of system, \sect{combine} presents details of the two-pass combination framework using $N$-best and lattice rescoring. The experimental setup and results are given in \twosect{setup}{exp}, and the paper concludes in \sect{conclusion}.
\section{Related Work}
\label{sec:related}
\subsection{Hybrid CTC and Attention-Based Models}
Watanabe \emph{et al.}~\cite{Watanabe2017HybridCA} proposed to augment AED with an extra output layer for CTC in the multi-task training framework. During decoding, beam search is performed based on the label-synchronous decoder of AED, and the frame-synchronous scores produced by the CTC output layer are interpolated with the label-synchronous scores of the current partial hypothesis to improve pruning~\cite{Watanabe2017HybridCA}. There are a few key differences between Watanabe's joint decoding approach~\cite{Watanabe2017HybridCA} and our proposed combination approach. First, our approach performs beam search in a frame-synchronous fashion. It is known that beam search with frame-synchronous systems can more easily handle streaming data and explore a larger search space. Second, although it is possible to incorporate the scores from two types of systems in a single decoding pass, our method achieves this using a frame-synchronous decoding pass followed by a separate rescoring pass by a label-synchronous system. The two-pass approach not only makes it easier to implement by reusing the existing LM rescoring framework, but also allows the two types of system to have different output units. Third, our method is configured to combine multiple systems and can use different model structures with or without sharing the encoder. This offers more flexibility in the choice of the systems and possibly more complementarity.

\subsection{Two-Pass End-to-End ASR}
Sainath \emph{et al.}~\cite{Sainath2019TwoPassES} proposed using an RNN-T to generate $N$-best hypotheses during the first pass and then rescore them using an AED, which is trained in the multi-task framework by sharing the encoders of RNN-T and AED. Using an adaptive beam and a prefix-tree representation for $N$-best list improves the rescoring results and speed. This is another instance of combining frame-synchronous and label-synchronous systems, which is probably the most similar method to ours. Compared to our approach, Sainath's approach~\cite{Sainath2019TwoPassES} pays more attention to on-device streaming constraints such as model size and latency, and therefore sacrifices the extra complementarity from different encoders. Furthermore, the use of the RNN-T decoder reduced the size of the search space to be explored due to its high computational cost, and thus only the $N$-best hypotheses are used for second pass rescoring. In contrast, our method allows a more flexible choice of the first pass frame-synchronous systems, including NN-HMMs and CTC in addition to RNN-T. Since NN-HMMs and CTC do not have a model-based decoder, it is easier for them to incorporate structured knowledge and construct richer lattices.

\subsection{Extending Our Previous Work: Integrating Source-Channel and Attention-Based Models}
This paper builds on the techniques and observations in our previous work~\cite{Li2019IntegratingSA}. In that work, we started by comparing Watanabe's joint decoding approach~\cite{Watanabe2017HybridCA} with our proposed two-pass combination approach, termed integrated source-channel and attention-based models (ISCA), under the multi-task training framework. In ISCA, both CTC and NN-HMMs were used as the source-channel model to perform the first pass decoding, and an RNN-based AED model was used to rescore the derived $N$-best hypotheses. It was shown that by using triphone targets together with a lexicon, ISCA outperformed the joint decoding approach. Further improvements were observed when multi-task training was not used, \textit{i.e.} the encoders of the source-channel and AED models are trained separately. Shortly afterwards, an approach similar to ISCA that combines frame-synchronous and label-synchronous systems at the hypothesis level was also studied~\cite{Wong2020CombinationOE}.

In this paper, the NN-HMM systems and AED models are built separately to allow them to have different model architectures, output units and training procedures, which maximises their complementarity and the combined performance. Furthermore, not only $N$-best but also lattice rescoring are investigated in this paper, and an improved lattice rescoring algorithm is proposed. For the AED models, Conformer encoders are used in this work instead of the LSTM encoders in \cite{Li2019IntegratingSA} for improved performance. Furthermore, this paper compares an RNN-based decoder with a Transformer-based decoder for $N$-best and lattice rescoring. Lastly, the combination of an NN-HMM system, an RNNLM, and two AED models with different decoders is studied. 

\section{Frame-Synchronous Systems}
\label{sec:fsync}
Speech recognition can be viewed as a \emph{noisy source-channel} modelling problem~\cite{jelinek1997statistical} where speech is produced and encoded via a noisy channel and the recogniser finds the most probable source text $\mathcal{W}^{*}$ given the observation sequence $\mathcal{O}$. That is,
\begin{equation}
    \mathcal{W}^{*}=\arg\max_{\mathcal{W}}P(\mathcal{W}|\mathcal{O}) \propto \arg\max_{\mathcal{W}}p(\mathcal{O}|\mathcal{W})P(\mathcal{W}),
    \label{eqn:scmodel}
\end{equation}
where $p(\mathcal{O}|\mathcal{W})$ is estimated by an AM and $P(\mathcal{W})$ is estimated by an LM \cite{jelinek1997statistical}. With standard HMM AMs, each decoding step consumes one acoustic frame, which makes it frame-synchronous.  

\subsection{NN-HMM Systems}
\label{ssec:hmm}
\begin{figure}[t]
    \centering
    \subfloat[]{\includegraphics[height=0.42in]{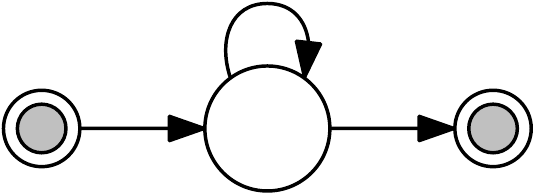}%
    \label{fig:hmm}}
    \hfil
    \subfloat[]{\includegraphics[height=0.5in]{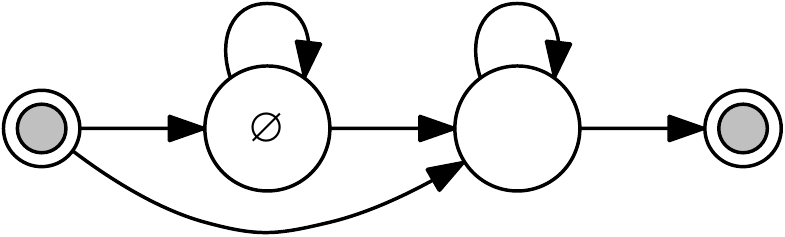}%
    \label{fig:ctc}}
    \caption{(a) is a single-state HMM. (b) is a CTC-equivalent HMM topology. The white and grey circles are emission and non-emission states, and $\varnothing$ is the blank symbol.}
    \label{fig:hmm&ctc}
\end{figure}

In acoustic modelling, HMMs (as shown in \fig{hmm}) together with NNs, are used to model the generative process of the observation sequences \textit{w.r.t.} the subword units. The model parameters are estimated by maximum likelihood (ML), whose loss function form can be written as
\begin{equation}
    \mathcal{L}_{\text{ML}}=-\ln \sum\nolimits_{\mathcal{S}}\prod\nolimits_{t=1}^{T}
    P(s_{t+1}|s_t)p(\mathbf{x}_t|s_t),
    \label{eqn:hmm}
\end{equation}
where $T$ is the length of the utterance; $\mathcal{S}$ is a collection of all possible HMM state sequences corresponding to the given reference transcription $\mathcal{W}^\text{ref}$, and $[s_{1:T}]\in\mathcal{S}$. $P(s_{t+1}|s_t)$ is the transition probability, which is less important compared to the $p(\mathbf{x}_t|s_t)$ for NN-HMMs. The observation probability $p(\mathbf{x}_t|s_t)$ can be estimated by
\begin{equation}
p(\mathbf{x}_t|s_t)\propto P(s_t|\mathbf{x}_t)/P(s_t),
\label{eqn:convert}
\end{equation}
where $\mathbf{x}_t$ is the input vector to an NN at time $t$, which is often obtained by stacking a subset of frames in 
$\mathcal{O}=[\mathbf{o}_{1:T}]$; $P(s_t|\mathbf{x}_t)$ is the output of the NN at $t$ relevant to $s_t$; $P(s_t)$ is the prior probability that can be estimated as the frequency of $s_t$ in the training data.

During training, the forward-backward procedure can be used to find the summation in Eqn.~(\ref{eqn:hmm})~\cite{Baum1967FwdBwd}. Alternatively, the summation can be approximated using only one state sequence found by performing state-to-frame alignment with a pre-trained ASR system~\cite{bourlard1994connectionist}. At test-time, by modifying \eqn{scmodel}, the AM and LM scores are combined based on a log-linear interpolation with an LM weight in decoding.

\subsection{CTC and RNN-T}
\label{ssec:ctc}
CTC trains an NN AM with a blank output symbol at the sequence-level without any explicit HMM structure~\cite{Graves2006CTC}. The training loss function is
\begin{align}
    \mathcal{L}_{\text{CTC}}=-\ln \sum\nolimits_{\mathcal{S}}\prod\nolimits_{t=1}^{T}P(s_t|\mathbf{x}_t),
    \label{eqn:ctc}
\end{align}
where $\mathcal{S}$ collects all possible symbol sequences that can map to $\mathcal{W}^\text{ref}$ by removing repeated symbols and blanks. During training, alignments and losses are computed by the forward-backward procedure.

By comparing \eqn{ctc} to \eqn{hmm}, CTC is equivalent to a special instantiation of the two-state HMM structure when $P(s_t)$ and ${P(s_{t+1}|s_t)}$ are constant for any state~\cite{Hadian2018EndtoendSR}. As shown in \fig{ctc}, the first emission state of the HMM is the skippable blank state with a self-loop and the second state corresponds to the subword unit. The blank state is shared across all HMMs. Therefore, it is reasonable to view CTC as a special HMM-based AM, which combines the HMM topology, the ML training loss and the forward-backward procedure. Although CTC is often used to model grapheme-based units to avoid using a complex decoder, CTC systems can have much better performance by re-introducing phone-based output units, lexicon and the frame-synchronous decoder~\cite{Li2019IntegratingSA}.

To remove the independence assumption across output tokens in CTC, RNN-transducer was proposed~\cite{Graves2012SequenceTW} where a prediction network and a joint network are added such that each non-blank output token depends on the previous ones. Nevertheless, since the underlying loss computation is the same as CTC and the decoding process runs frame-by-frame, RNN-Ts are also frame-synchronous. Recent work has shown that, with a limited history used for the prediction network, RNN-Ts can be decoded similarly as the HMM-based AM with a lexicon and an LM~\cite{Variani2020HybridAT}.

\section{Label-Synchronous Systems}
\label{sec:lsync}
An AED model maps a $T$-length input sequence $\mathcal{O}$ to an $L$-length output subword sequence $\mathcal{C}=[c_{1:L}]$ where the input length is normally longer than the output length. Instead of decomposing into AMs and LMs as in \sect{fsync}, the AED model computes the posterior distribution $P(\mathcal{C}|\mathcal{O})$ directly following the chain rule of conditional probability:
\begin{equation}
 P(\mathcal{C}|\mathcal{O})=P(c_1|\mathcal{O})\prod\nolimits_{l=2}^{L}P(c_l|c_{1:l-1},\mathcal{O}),
    \label{eqn:seqprob}
\end{equation}
The encoder extracts acoustic features while the decoder generates hypotheses, and the acoustic and language information is jointly learned using a single model without making any independence assumption. The neural decoder processes label-by-label, and $P(\mathcal{C}|\mathcal{O})$ is maximised using the per-label cross-entropy loss in training. Hence, AED models are \emph{label-synchronous}.

\subsection{Neural Encoder}
The neural encoder of an AED model extracts a sequence of hidden representations $\mathcal{E}$ from $\mathcal{O}$ using an NN architecture,
\begin{equation}
    \mathcal{E} = \text{NeuralEncoder}(\mathcal{O}).
\end{equation}
Commonly used NN architectures for $\text{NeuralEncoder}$ include a stack of RNN layers, such as long short-term memory (LSTM) layers ~\cite{Chorowski2015AttentionBasedMF,Chan2016ListenAA}, Transformer encoder blocks~\cite{Vaswani2017AttentionIA}, and Conformer encoder blocks~\cite{Gulati2020ConformerCT}.

\subsection{Neural Decoder}
Label-synchronous systems can be categorised into two types according to the decoder architecture.
\begin{figure}[t]
    \centering
    \subfloat[]{\includegraphics[width=.535\linewidth]{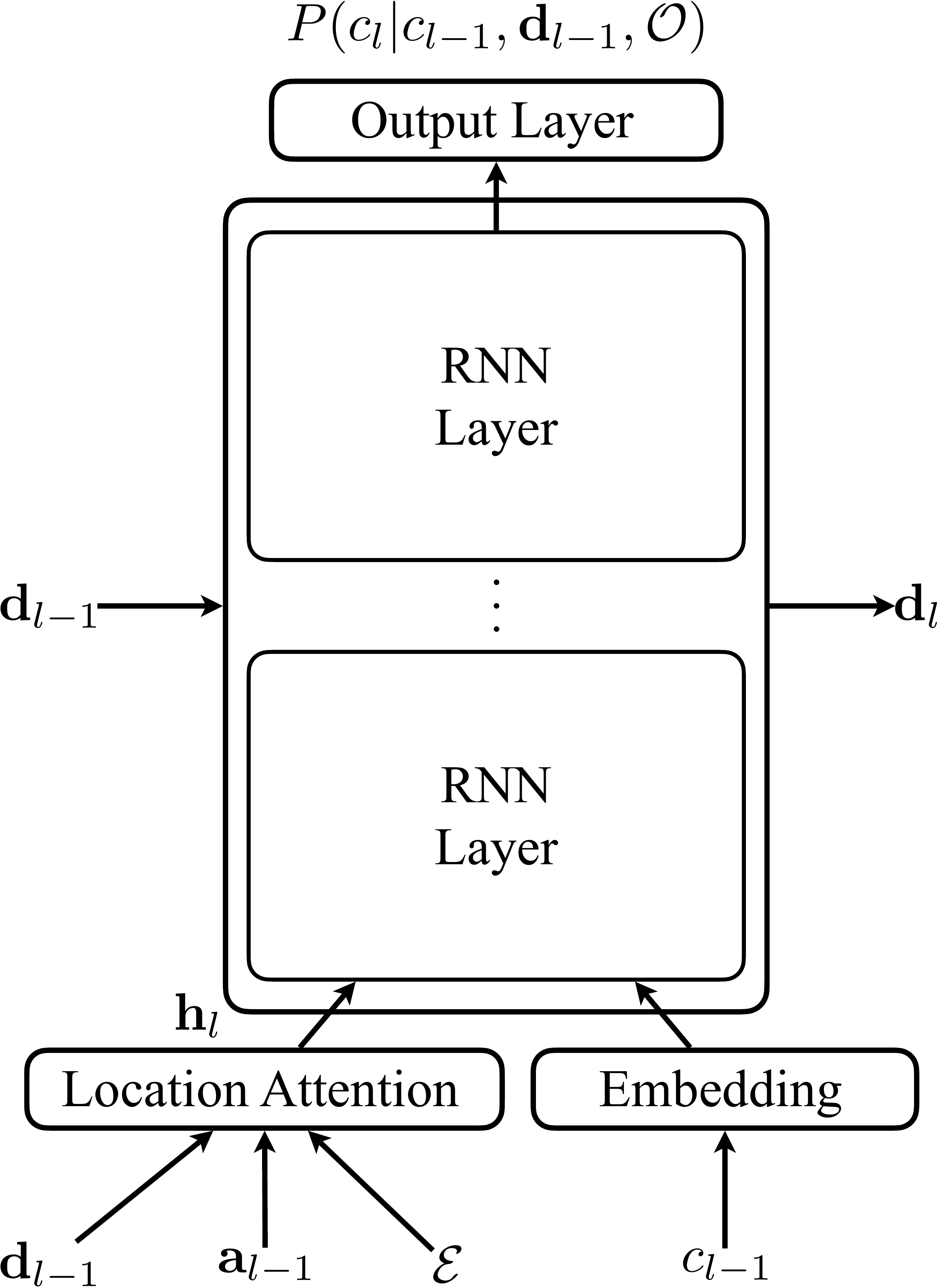}%
    \label{fig:rnn_dec}}
    \hfil
    \subfloat[]{\includegraphics[width=0.46\linewidth]{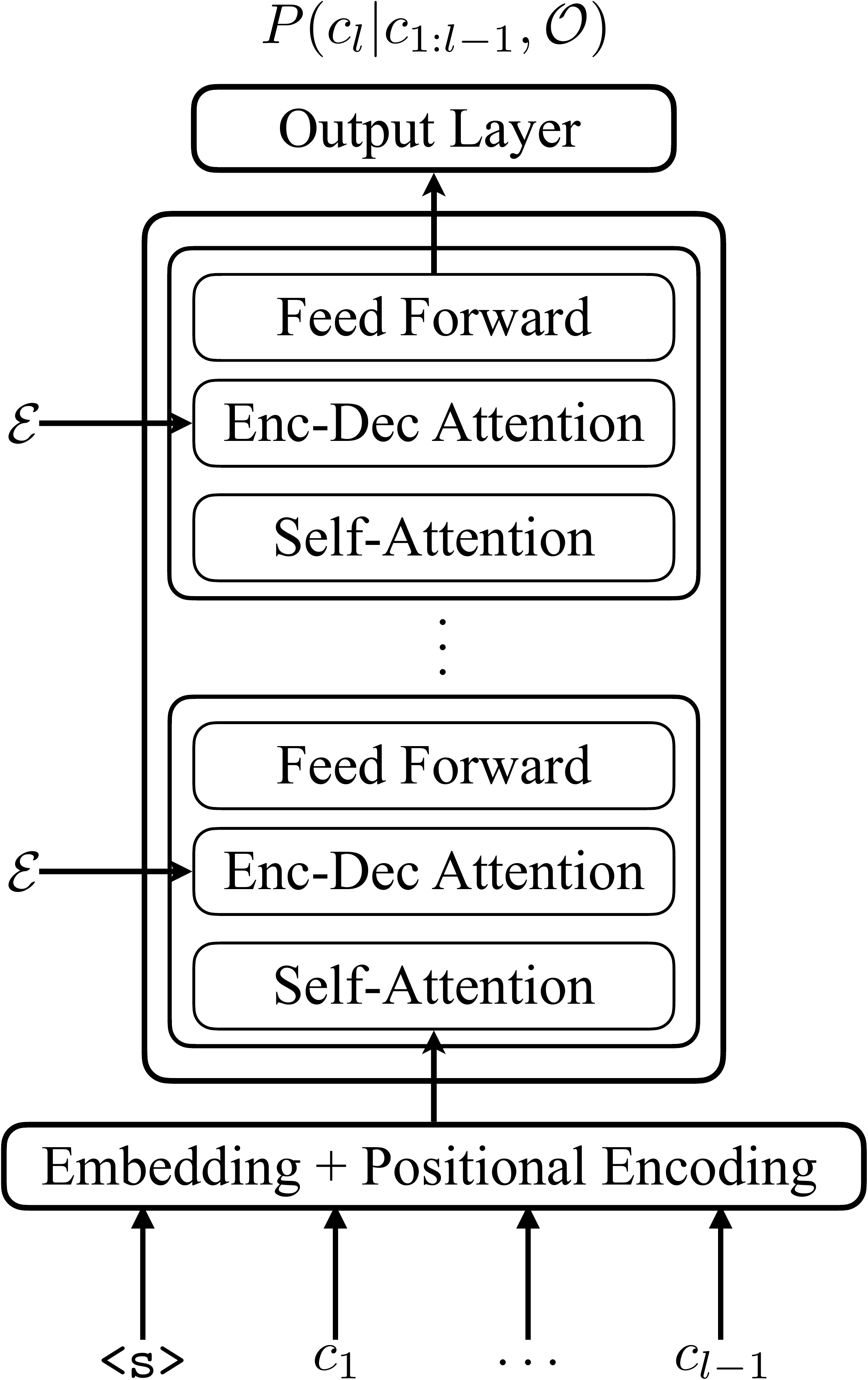}%
    \label{fig:tfm_dec}}
    \caption{(a) An RNN neural decoder and (b) a Transformer neural decoder predicting the output distribution of subword unit $c_l$ at the $l$-th step.}
    \label{fig:decs}
\end{figure}
\subsubsection{RNN Decoder}
The first generation of AED models were RNN-based~\cite{Chorowski2015AttentionBasedMF,Chan2016ListenAA}. As shown in \fig{rnn_dec}, at each decoding step $l$, $\mathcal{E}$ is transformed into a context vector $\mathbf{h}_{l}$ based on the annotation vector $\mathbf{a}_{l}$ produced by the attention mechanism. The RNN decoder is assumed to satisfy the first-order Markov property~\cite{Bengio1993}, which assumes $P(c_l|c_{1:l-1},\mathcal{O})= P(c_l|c_{l-1},\mathbf{d}_{l-1},\mathcal{O})$ by conditioning on $\mathcal{O}$ and the previous decoder state $\mathbf{d}_{l-1}$. The decoding procedure stops when the end-of-sentence symbol is generated, which allows output sequences to have variable lengths. More specifically,
\begin{align}
    \mathbf{a}_l & = \text{Attention}(\mathbf{a}_{l-1},\mathbf{d}_{l-1},\mathcal{E})\\
    \mathbf{h}_l & = \mathcal{E}\;\mathbf{a}_{l}\\
    P(c_l|c_{l-1},\mathbf{d}_{l-1},\mathcal{O}),\mathbf{d}_{l} & = \text{RNNDecoder}(c_{l-1},\mathbf{d}_{l-1},\mathbf{h}_l).\label{eqn:neuraldecoder}
\end{align}

\subsubsection{Transformer Decoder}
Transformers~\cite{Vaswani2017AttentionIA} have recently emerged as another type of AED model without recurrent structures. Instead of the location-based attention-mechanism used by RNN-based AED, Transformers use multi-head scaled dot-product attention and positional encoding to process all time steps in a sequence in parallel while preserving the sequential information, which greatly accelerates training. As illustrated in \fig{tfm_dec}, a Transformer decoder can have multiple blocks, where each block consists of a self-attention layer that performs multi-head attention over the decoder input or the output from the previous decoder block, an encoder-decoder attention layer that performs multi-head attention over the encoder output and a feed-forward module. A Transformer decoder is denoted as 
\begin{align}
    P(c_l|c_{1:l-1},\mathcal{O}) & = \text{TransformerDecoder}(c_{1:l-1}, \mathcal{E}).
\end{align}
At test-time, a sequential step-by-step decoding procedure is still necessary for Transformer decoders. Meanwhile, unlike RNN decoders that employ the first-order Markov property, the hidden representations produced by all blocks in the Transformer decoder at all previous time steps are required to be stored throughout the entire decoding procedure for self-attention, which makes it less memory efficient. \tbl{complexity} shows the complexities of different decoders, which indicates that the RNN decoder is more suitable for long utterances due to its high memory efficiency while Transformer decoders may perform better since they do not make the Markov assumption. 

\begin{table}[t]
    \caption{Comparison of the Complexities of RNN and Transformer Decoders with Respect to Output Sequence Length $L$.}
    \centering
    \begin{tabular}{cccccc}
        \toprule
        \multirow{2}{*}{neural decoder} & \multicolumn{2}{c}{computation complexity} & & \multicolumn{2}{c}{storage complexity}\\
        \cmidrule{2-3}\cmidrule{5-6}
         &  training & test & & training & test \\
        \midrule
        RNN & $O(L)$ & $O(L)$ & & $O(L)$ & $O(1)$\\
        Transformer & $O(1)$ & $O(L)$ & & $O(L^2)$ & $O(L^2)$\\
        \bottomrule
    \end{tabular}
    \label{tab:complexity}
\end{table}

\section{Combining Frame-Synchronous and Label-Synchronous Systems}
\label{sec:combine}
\begin{figure*}[t]
    \centering
    \includegraphics[width=0.7\linewidth]{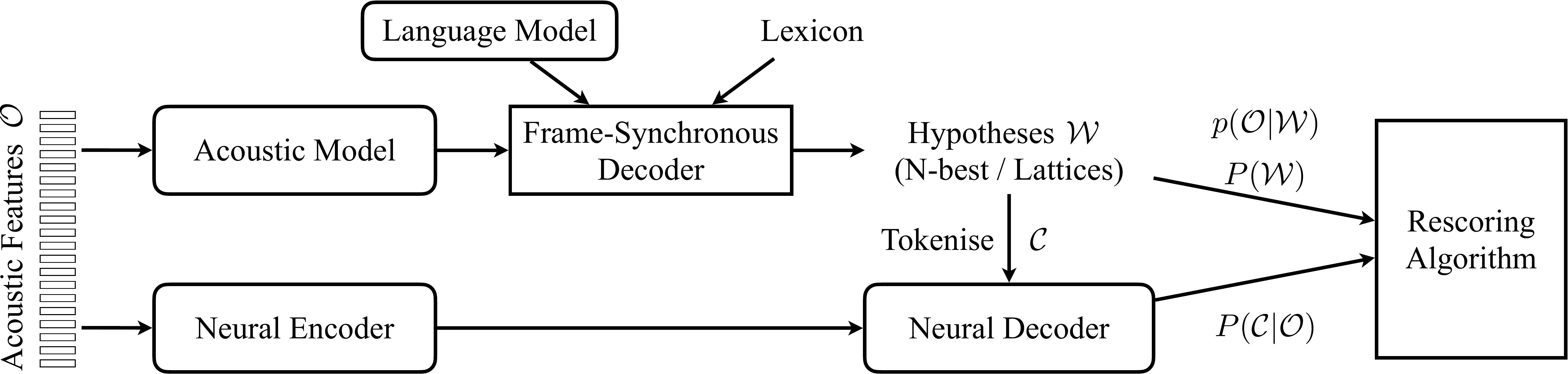}
    \caption{Pipeline for combining a frame-synchronous system and a label-synchronous system. Rounded box indicates trainable models and rectangular box indicates an algorithm or a procedure.}
    \label{fig:isca}
\end{figure*}
\subsection{Motivation}
\label{ssec:motivation}
Frame-synchronous and label-synchronous systems exhibit distinctive advantages due to their different modelling and decoding strategies. Frame-synchronous systems such as NN-HMMs and CTC assume the calculations of different $P(s_t|\mathbf{x}_t)$ are independent. In contrast, label-synchronous systems such as AED models calculates $P(\mathcal{C}|\mathcal{O})$ based on \eqn{seqprob} without making any label independence assumption. Consequently, frame-synchronous systems with phone-based acoustic models can easily incorporate structured phonetic, linguistic and contextual knowledge by using a lexicon. This is very helpful for the robustness of the ASR systems such as rare words. Because of the label independence assumption, NN-HMMs and CTC can produce a rich and compact graph-based representation of the hypothesis space for each utterance such as a lattice and a confusion network. Recent studies show that by limiting the history context, \emph{i.e.} imposing some degree of label independence assumption, RNN-Ts can also effectively generate lattices with comparable performance to the full-context RNN-Ts~\cite{Prabhavalkar2020LessIM}. It is more difficult for label-synchronous systems to generate lattices since the previous decoding output needs to be fed into the auto-regressive neural decoder to obtain the probability distribution over the next token. The hypothesis space explored by label-synchronous systems is normally limited to the top $N$ candidates, which is significantly less than lattices from frame-synchronous systems. Although it is more challenging to adapt AED models to process streaming data due to its label-synchronous nature~\cite{Miao2020OnlineHC,Moritz2020StreamingAS}, AED models can jointly model acoustic and textual information and the neural decoder can model long-range dependencies across labels without the independence assumption. 

Therefore, it is desirable to leverage the advantages of both types of systems and exploit their complementarity to improve the performance of the final ASR system. In this work, we propose a generic two-pass ASR combination method, which uses the frame-synchronous systems to perform the first-pass decoding and uses the label-synchronous system to rescore the first-pass results. In the first-pass decoding with NN-HMMs or CTC systems, the model can process streaming data, leverage structured knowledge in particular contextual knowledge, and recover from search errors made in early decoding steps. In the second-pass rescoring with AED systems, complete acoustic and language information from each entire utterance can be used jointly to refine the estimation of hypothesis probabilities. Furthermore, since NN-HMM or CTC systems prune most of the less possible hypotheses, rescoring with AED can be more robust and efficient than decoding with AED models. 

\subsection{Combination Framework}
\label{ssec:framework}
As shown in \fig{isca}, the acoustic features of an utterance are first passed through the acoustic model of the frame-synchronous system. Together with a language model and a lexicon, the frame-synchronous decoder generates the hypotheses in the form of an $N$-best list or a lattice with corresponding acoustic and language model scores. The same acoustic features can be forwarded through the neural encoder of the label-synchronous system. The neural decoder then uses the hidden representations of the acoustic signal and scores the hypotheses from the first-pass similar to the training mode. Finally, applying either $N$-best rescoring or lattice rescoring with tuned interpolation coefficients yields the best hypothesis of the combined system. Generically, if a frame-synchronous system $P_f$ and a label-synchronous system $P_l$ are available, the final score of a hypothesis $\mathcal{W}$ for an utterance $\mathcal{O}$ is
\begin{align}
    \mathcal{S}(\mathcal{W}|\mathcal{O}) = \log P_f(\mathcal{W}|\mathcal{O}) + \alpha \log P_l(\mathcal{W}|\mathcal{O})
\end{align}
where $\alpha$ is the interpolation coefficient.

For HMM-based frame-synchronous systems, AM scores and $n$-gram LM scores are available as in \eqn{scmodel}. When combining with label-synchronous systems, the LM scaling factor can also be tuned together with other interpolation coefficients. For example, when rescoring with an RNNLM $P_{\text{RNNLM}}$, a label-synchronous system with an RNN decoder $P_{l-\text{RNN}}$, and a label-synchronous system with a Transformer decoder $P_{l-\text{TFM}}$, the final score becomes
\begin{align}
    \mathcal{S}(\mathcal{W}|\mathcal{O}) = & \log p(\mathcal{O}|\mathcal{W}) \nonumber\\
    & + \gamma \log P_{\text{$n$-gram}}(\mathcal{W}) + \lambda\log P_{\text{RNNLM}}(\mathcal{W}) \nonumber\\
    & + \alpha\log P_{\text{RNN}}(\mathcal{W}|\mathcal{O}) + \beta\log P_{\text{TFM}}(\mathcal{W}|\mathcal{O})\nonumber \\
    & + \kappa |\mathcal{W}|
    \label{eqn:score}
\end{align}
where $\gamma,\lambda$ are the coefficients for two LMs, $\alpha,\beta$ are the coefficients for two label-synchronous systems, $\kappa$ is the insertion penalty, and $|\mathcal{W}|$ is the number of words in the hypothesis. Based on \eqn{score}, label-synchronous systems can be viewed as audio-grounded language models. For other frame-synchronous systems, corresponding scores from various systems can be interpolated in a similar fashion.

In contrast to the proposed framework, standard combination methods for ASR systems are not suitable for combining with label-synchronous systems. For example, confusion network combination~\cite{Evermann2000PosteriorPD} requires decoding lattices and ROVER~\cite{Fiscus1997APS} requires comparable confidence measures from both types of system. The benefits of using hypothesis-level combination may be limited, because the hypothesis space generated from the frame-synchronous system is generally much larger than the label-synchronous system.

\begin{algorithm*}[t]
    \caption{Lattice Rescoring Using a Label-Synchronous Model\protect\footnotemark}
    \label{alg:latrescore}
    \begin{algorithmic}[1]
    \Procedure{LatticeRescore}{lattice, model, utt, ngram, collar}
        \State cache$\gets$ \textsc{CreateCache}(model, utt, collar) \Comment{Initialise the two-level cache}
        \For{node $n_i$ in lattice.nodes} \Comment{Initialise expanded nodes and arcs}
            \State $n_i$.expanded\_nodes $\gets$ []; $n_i$.expanded\_arcs $\gets$ []
        \EndFor
        \State lattice.nodes[0].expanded\_nodes.Append(lattice.nodes[0].Duplicate()) \Comment{Initialise the starting node}
        \For{node $n_i$ in lattice.nodes} \Comment{Lattice traversal, expansion and rescoring}
            \For{node $\tilde{n}_j$ in $n_i$.expanded\_nodes}
                \For{arc $a_k$ in $n_i$.exits}
                    \State $n_k$ $\gets$ $a_k$.dest
                    \State hist $\gets$ \textsc{LastNItems}([$\tilde{n}_j$.hist, $n_k$.word], ngram-1)
                    \If{$\exists$ node $\tilde{n}_l\in n_k$.expanded\_nodes such that $\tilde{n}_l$.hist = hist}
                        \State $\tilde{a}_l$ $\gets$ $a_k$.Duplicate($\tilde{n}_j$, $\tilde{n}_l$) \Comment{Create arc between expanded nodes}
                    \Else
                        \State $\tilde{n}_l$ $\gets$ $n_k$.Duplicate(hist) \Comment{Create expanded node with new history}
                        \State $n_k$.expanded\_nodes.Append($\tilde{n}_l$)
                        \State $\tilde{a}_l$ $\gets$ $a_k$.Duplicate($\tilde{n}_j$, $\tilde{n}_l$) \Comment{Create arc between expanded nodes}
                    \EndIf
                    \State post $\gets$ \textsc{LastNItems}([cache.GetPost($\tilde{n}_j$.hist, $\tilde{n}_j$.time), $\tilde{a}_l$.post], ngram-1)
                    \If{cache.Lookup(hist, $n_k$.time) fails}
                        \State cache.Renew($\tilde{n}_j$.hist, $\tilde{n}_j$.time, $\tilde{n}_l$.hist, $\tilde{n}_l$.time, post) \Comment{Cache miss -- new ngram or timestamp}
                    \ElsIf{\textsc{Sum}(post) $>$ \textsc{Sum}(cache.GetPost($\tilde{n}_l$.hist, $\tilde{n}_l$.time))}
                        \State cache.Renew($\tilde{n}_j$.hist, $\tilde{n}_j$.time, $\tilde{n}_l$.hist, $\tilde{n}_l$.time, post) \Comment{Cache update -- use the more likely path}
                    \EndIf
                    \State $\tilde{a}_l$.model\_score $\gets$ cache.GetPred($\tilde{n}_j$.hist, $\tilde{n}_j$.time, $\tilde{n}_l$.word) \Comment{Assign score from label-sync model}
                    \State $\tilde{n}_j$.exits.Append($\tilde{a}_l$); $\tilde{n}_l$.entries.Append($\tilde{a}_l$) \Comment{Connect new arc to expanded nodes}
                \EndFor
            \EndFor
        \EndFor
        \State lattice $\gets$ \textsc{BuildExpandedLattice}(lattice) \Comment{Construct new lattice from expanded nodes and arcs}
    \EndProcedure
    \end{algorithmic}
\end{algorithm*}

\subsection{Training}
\label{ssec:training}
The frame-synchronous system and the label synchronous system can be trained separately or jointly in a multi-task fashion by sharing the neural encoder with the acoustic model. For multi-task trained models~\cite{Watanabe2017HybridCA}, the total number of parameters in the entire system is smaller due to parameter sharing. Although multi-task training can be an effective way of regularisation, setting the interpolation weights between the two losses and configuring the learning rate to achieve good performance for both models may not be straightforward. Moreover, multi-task training also limits the model architectures or model-specific training techniques that can be adopted for individual systems. For example, the acoustic model can be a unidirectional architecture for streaming purposes but the neural encoder for the label-synchronous system can be bi-directional for second-pass rescoring. Acoustic models in frame-synchronous systems normally have a frame subsampling rate of 3 in a low frame-rate system~\cite{Pundak2016LowerFR,Povey2018SemiOrthogonalLM} but neural encoders normally have a frame rate reduction of 4, by using convolutional layers~\cite{Gulati2020ConformerCT} or pyramidal RNNs~\cite{Chan2016ListenAA}, for better performance. Frame-level shuffling~\cite{Su2013ErrorBP} is important for the optimisation of HMM-based acoustic models, whereas label-synchronous systems have to be trained on a per utterance basis. Triphone units are commonly used for HMM-based acoustic models whereas word-pieces are widely used for attention-based models~\cite{Chiu2018StateoftheArtSR}. Overall, sharing the acoustic model and the neural encoder in a multi-task training framework hampers both systems from reaching the best possible performance. Therefore, this work will focus on different systems that are trained separately.

\subsection{N-best Rescoring}
\label{ssec:nbest}
The frame-synchronous system generates the top $N$ hypotheses for each utterance. The word sequence $\mathcal{W}$ can be tokenised into the set of word-pieces modelled by the label-synchronous system. The word-piece sequence $\mathcal{C}$ can be forwarded through the neural decoder to obtain the probability for each token $P(c_t|c_{1:t-1},\mathcal{O})$. By tuning the interpolation coefficients in \eqn{score} to have the lowest WER on the development set, the final hypothesis is the one with the highest score among the top $n$ candidates.
\begin{equation}
    \mathcal{W}^{*} = \argmax_{\mathcal{W}} \mathcal{S}(\mathcal{W}|\mathcal{O})
\end{equation}
When rescoring the $N$-best hypotheses with a label-synchronous system with an RNN-based decoder, the time complexity is $O(L)$ and the space complexity is $O(1)$ because of the sequential nature of RNNs. In contrast, a Transformer-based decoder has a time complexity of $O(1)$ and space complexity of $O(L^2)$ as during Transformer training. Since the entire hypothesis is available, self-attention can be directly computed across the whole sequence for each token.

\subsection{Improved Lattice Rescoring}
\label{ssec:lattice}
\begin{figure}[t]
    \centering
    \includegraphics[width=0.9\linewidth]{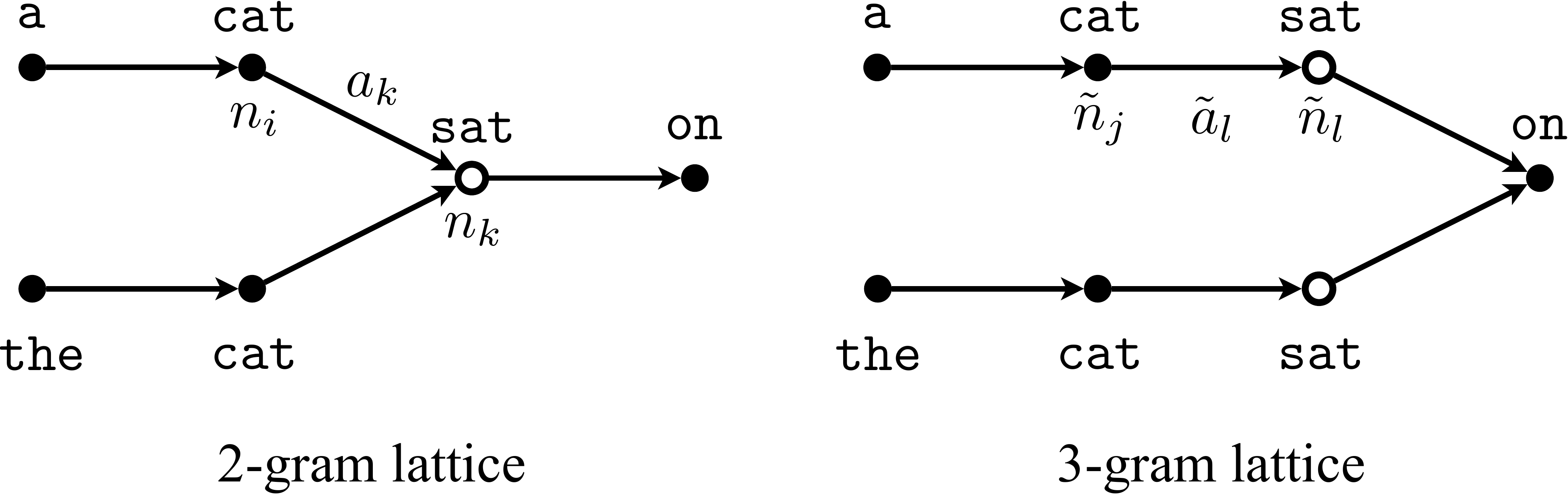}
    \caption{Example of a 2-gram lattice expanding to a 3-gram lattice. The hollow node from the left is expanded into two hollow nodes on the right, which corresponds to line 16-18 in \alg{latrescore}.}
    \label{fig:latexp}
\end{figure}
\footnotetext{Our code is available at \url{https://github.com/qiujiali/lattice-rescore}.}
For a fixed number of candidate hypotheses, the number of alternatives per word in the sequence is smaller when the hypothesis becomes longer. This means the potential for improvement diminishes for longer utterances. Therefore, in order to have the same number of alternatives per word, the size of $N$-best lists needs to grow exponentially with respect to the utterance length. However, lattice rescoring can effectively mitigate this issue. Lattices are directed acyclic graphs where nodes represent words and edges represent associated acoustic and language model scores. A complete path from the start to the end of a lattice is a hypothesis. Because a various number of arcs can merge to or split from a node, a lattice generally contains a far greater number of hypotheses than a limited $N$-best list. The size of lattices is measured by the number of arcs per second of speech, also known as the \emph{lattice density}.

One commonly used lattice rescoring approach for RNNLMs is on-the-fly lattice expansion with $n$-gram based history clustering~\cite{Liu2016TwoEL}. Similar to RNNLMs, attention-based models are auto-regressive models where the current prediction depend on all history tokens. This means approximations must be made when assigning scores on edges of lattices because each word in the lattice may have numerous history sequences. Although lattices can be expanded to allow each word to have a more unique history, a trade-off between the uniqueness of the history and computational efficiency need to be considered. $n$-gram based history clustering~\cite{Liu2016TwoEL} assumes that the history before the previous $n-1$ words has little impact on the probability of the current word. As illustrated in \fig{latexp}, a lattice can be expanded such that the $n-1$ history words of each word in the lattice are unique. During rescoring, a hash table-based cache is created, where the key is the $n-1$ history words and the value is the corresponding hidden state of RNNLM and the output distribution. When the same history appears again during rescoring, repetitive computation can be avoided, regardless of the more distant history. 

However, for label-synchronous systems based on attention mechanisms, $n$-gram based history clustering can lead to undesirable behaviour. For example, ``\texttt{i think those are are wonderful things to have but i think in a big company}'' (from SwitchBoard), the phrase ``\texttt{i think}'' appears twice in the utterance. If the original trigram based history clustering is used, at the second occurrence of the phrase, the algorithm will restore the cache from the first occurrence including the attention context and the decoder state from the first occurrence and then continue to score the rest of the utterance. Consequently, scores for the second half of the utterance will be wrong because of the incorrect attention context. To this end, a time-dependent two-level $n$-gram cache is proposed. When looking up the history phrase in the cache, a secondary level cache indexed by the corresponding frame number of the word is used. 

A collar of $\pm9$ frames is used when looking up the secondary cache to accommodate a small difference in alignment. More specifically, when looking up the cache, the timestamp of the current word is also used. The cache is only hit when the timestamp falls within the vicinity of one of the timestamps in the cache. Otherwise, the cache is missed and a new entry is created in the cache with the timestamp and the corresponding decoder states (line 22 in \alg{latrescore}). For all the functions related to the cache, a timestamp must be provided. Another key detail is that when there is a cache hit, the sum of arc posteriors of the current $(n-1)$-gram is compared with the one stored in the cache. If the current posterior is larger, indicating the current $(n-1)$-gram is on a better path, then the cache entry is updated to store the current hidden states (line 24 in \alg{latrescore}). For the example in \fig{latexp}, when the lower path is visited after the upper path, the cache entry ``\texttt{sat on}'' should already exist. If the lower path has a higher posterior probability, then the cache entry will be updated, so that future words in the lattice will adopt the history from the lower path.

For label-synchronous systems with RNN decoders, lattice rescoring has $O(L)$ for time complexity and $O(1)$ for space complexity as the RNN hidden states can be stored and carried forward at each node in the lattice. However, since lattice rescoring operates on partial hypotheses, Transformer decoders have to run in the decoding mode as in \tbl{complexity}. Because self-attention need to be computed with all previous tokens, lattice rescoring with the Transformer-based decoder has $O(L^2)$ time and space complexities.

\section{Experimental Setup}
\label{sec:setup}
\subsection{Acoustic Data}
Two common ASR benchmarks are used for training and evaluation. The Augmented multi-party interaction (AMI)~\cite{Carletta2005TheAM} dataset is relatively small-scale and contains recordings of spontaneous meetings. The individual headset microphone (IHM) channel is used. As shown in \tbl{data}, it has many short utterances as multiple speakers take short turns during meetings. The original corpus has official development (dev) and evaluation (eval) sets. Switchboard-1 Release 2 (SWB-300) is a larger-scale dataset with telephony conversations. Compared to AMI, it has more training data, longer utterances, more speakers and a larger vocabulary. Hub5'00 is used as the development set, which is split into two subsets: Switchboard (SWB) and CallHome (CHM). Note that the SWB subset has overlapping speakers with the SWB-300 training set. RT03 is used as the evaluation set, which is split into Switchboard Cellular (SWBC) and Fisher (FSH) subsets\footnote{The LDC catalogue numbers are LDC97S62 for SWB-300, LDC2002S09 and LDC2002T43 for Hub5'00, and LDC2007S10 for RT03.}. The acoustic data preparation follows the Kaldi recipes~\cite{Povey2011TheKS}.
\begin{table}[t]
    \caption{Two Datasets Used for Experiments.}
    \centering
    \begin{tabular}{lcc}
        \toprule
         & AMI-IHM & SWB-300 \\
        \midrule
        style & meeting & telephony \\
        training data & 78 hours & 319 hours\\
        avg. utterance length & 7.4 words & 11.8 words\\
        number of speakers & 155 & 520 \\
        vocabulary size & 12k & 30k \\
        \bottomrule
    \end{tabular}
    \label{tab:data}
\end{table}

\subsection{Text Data and Language Models}
\begin{table}[t]
    \caption{Perplexities of Various Language Models on Two Datasets. Dev and Eval Sets for SWB-300 are Hub5'00 and RT03.}
    \centering
    \begin{tabular}{lccc}
        \toprule
        dataset & LM & dev & eval\\
        \midrule
        \multirow{3}{*}{AMI-IHM} & 3-gram & 80.2 & 76.7\\
        & 4-gram & 79.3 & 75.7\\
        & RNNLM & 58.0 & 53.5\\
        \midrule
        \multirow{3}{*}{SWB-300} & 3-gram & 82.8 & 67.7\\
        & 4-gram & 80.0 & 65.3\\
        & RNNLM & 51.9 & 45.3\\
        \bottomrule
    \end{tabular}
    \label{tab:ppl}
\end{table}
For each dataset, its training transcription and Fisher transcription\footnote{The LDC catalogue numbers are LDC2004T19 and LDC2005T19 for Fisher transcription.} are used to train both $n$-gram language models and RNNLMs. Text processing and building $n$-gram LMs for both datasets also follow the Kaldi recipes~\cite{Povey2011TheKS}. RNNLMs are trained using the ESPnet toolkit~\cite{watanabe2018espnet}. The vocabulary used for RNNLMs is the same as for $n$-gram LMs and has 49k words for AMI-IHM and 30k words for SWB-300. RNNLMs have 2-layer LSTMs with 2048 units in each layer. Models are trained with stochastic gradient descent (SGD) with a learning rate of 10.0 and a dropout rate of 0.5. The embedding dimension is 256. Gradient norms are clipped to 0.25 and weight decay is set to $10^{-6}$. Training transcription and Fisher transcription are mixed in a 3:1 ratio. Because of the domain mismatch between AMI and Fisher text data, the RNNLM for AMI is fine-tuned on AMI transcriptions after training using the mixture of data with a learning rate of 1.0. The AMI RNNLM has 161M parameters and the Switchboard RNNLM has 122M parameters. The perplexities of LMs for both datasets are in \tbl{ppl}.

\subsection{Acoustic Models and Label-Synchronous Systems}
\begin{table}[t]
    \caption{Single System WERs on AMI-IHM and SWB-300 Datasets. Systems Do Not Use RNNLMs for Rescoring or Decoding.}
    \centering
    \begin{tabular}{lccccc}
        \toprule
         & \multicolumn{2}{c}{AMI-IHM} & & \multicolumn{2}{c}{SWB-300} \\ \cmidrule{2-3}\cmidrule{5-6}
         & \multirow{2}{*}{dev} & \multirow{2}{*}{eval} & & Hub5'00 & RT03 \\
         & & & & (SWB/CHM) & (SWBC/FSH) \\
        \midrule
        frame-sync & 19.9 & 19.2 & & 8.6 / 17.0 & 18.8 / 11.4\\
        label-sync-LSTM & 19.6 & 18.2 & &7.5 / 15.3 & 16.2 / 10.7\\
        label-sync-TFM & 19.4 & 19.1 & & 7.8 / 14.4 & 17.5 / 10.4\\
        \bottomrule
    \end{tabular}
    \label{tab:single_wer}
\end{table}
Acoustic models of the frame-synchronous systems are factorised TDNNs trained with lattice-free maximum mutual information objective~\cite{Povey2018SemiOrthogonalLM} by following the standard Kaldi recipes~\cite{Povey2011TheKS}. The total numbers of parameters are 10M for AMI-IHM and 19M for SWB-300.

Two types of label-synchronous system are trained using the ESPnet toolkit~\cite{watanabe2018espnet} without the CTC branch. Neural encoders are composed of 2 convolutional layers that reduce the frame rate by 4, followed by 16 Conformer blocks~\cite{Gulati2020ConformerCT}. For the Conformer block, the dimension for the feed-forward layer is 2048, the attention dimension is 512 for the model with an RNN decoder and 256 for the model with a Transformer decoder. The number of attention heads is 4 and the convolutional kernel size is 31. For the label-synchronous system with an RNN decoder (label-sync-LSTM), the decoder has a location-aware attention mechanism and 2-layer LSTMs with 1024 units. For the label-synchronous system with the Transformer-based decoder (label-sync-TFM), the decoder has 6-layer Transformer decoder blocks where the attention dimension is 256 and the feed-forward dimension is 2048. Both the label-sync-LSTM and label-sync-TFM are trained using the Noam learning rate scheduler on the Adam optimiser. The learning rate is 5.0 and the number of warmup steps is 25k. Label smoothing of 0.1 and a dropout rate of 0.1 are applied during training. An exponential moving average of all model parameters with a decay factor of 0.999 is used. The label-sync-LSTM has 130M parameters while the label-sync-TFM has 54M parameters. Beam search with a beam-width of 8 is used for decoding. Apart from applying length normalisation for the label-sync-TFM model, other decoding heuristics are not used. SpecAugment~\cite{Park2019SpecAugmentAS} and speed perturbation are applied. Word-piece outputs~\cite{Kudo2018SubwordRI} are used with 200 units for AMI-IHM and 800 units for SWB-300.

The single model WERs for the frame-synchronous system and label-synchronous systems are given in \tbl{single_wer}.
\section{Experimental Results}
\label{sec:exp}
\begin{table*}[t]
    \caption{WERs on AMI-IHM Eval Set Using $N$-Best and Lattice Rescoring With Various Combination of RNNLM and Two Label-Synchronous Systems. Lattice Density (Number of Arcs per Second) Is Provided in Square Brackets.}
    \centering
    \begin{tabular}{cccccccccc}
        \toprule
        \multirow{2}{*}{RNNLM} & \multirow{2}{*}{label-sync-LSTM} & \multirow{2}{*}{label-sync-TFM} && 20-best & 100-best & 500-best && 4-gram lattice & 5-gram lattice\\
        & & && [34.4] & [146.9] & [595.2] && [313.3] & [606.0]\\
        \midrule
        \checkmark & & && 16.9 & 16.5 & 16.3 && 16.3 & 16.3 \\
        & \checkmark & && 16.2 & 15.8 & 15.5 && 15.3 & 15.3\\
        & & \checkmark && 16.3 & 15.7 & 15.4 && 15.3 & 15.3\\
        \midrule
        \checkmark & \checkmark & && 15.4 & 14.7 & 14.3 && 14.1 & 14.1\\
        \checkmark & & \checkmark && 15.5 & 14.8 & 14.3 && 14.2 & 14.1\\
        \midrule
        \checkmark & \checkmark & \checkmark && 15.3 & 14.5 & 14.1 && \textbf{13.7} & 13.8\\
        \bottomrule
    \end{tabular}
    \label{tab:ami_combine}
\end{table*}
\begin{table*}[t]
    \caption{WERs on Hub5'00 and RT03 Using 500-Best Rescoring and Lattice Rescoring (5-gram Approximation) With Various Combination of RNNLM and Two Label-Synchronous Systems. Lattice Density (Number of Arcs per Second) Is Provided in Square Brackets.}
    \centering
    \begin{tabular}{ccccccccc}
        \toprule
        \multirow{2}{*}{RNNLM} &\multirow{2}{*}{label-sync-LSTM} & \multirow{2}{*}{label-sync-TFM} && \multicolumn{2}{c}{Hub5'00 (SWB/CHM)} & & \multicolumn{2}{c}{RT03 (SWBC/FSH)}\\[-0.2em]
        \cmidrule{5-6}\cmidrule{8-9}
        & & && 500-best [760.7] & 5-gram lattice [406.8] & & 500-best [721.2] & 5-gram lattice [496.2] \\
        \midrule
        \checkmark & & && 6.8 / 14.3 & 6.8 / 14.7 && 16.1 / 9.4 & 16.2 / 9.6 \\
        & \checkmark & && 6.5 / 13.3 & 6.5 / 13.2 && 15.2 / 9.0 & 14.7 / 8.8 \\
        & & \checkmark && 6.4 / 12.9 & 6.3 / 12.7 && 14.9 / 8.6 & 14.6 / 8.4\\
        \midrule
        \checkmark & \checkmark & && 5.9 / 12.7 & 5.8 / 12.7 && 14.3 / 8.2 & 13.8 / 7.9\\
        \checkmark & & \checkmark && 5.8 / 12.6 & 5.9 / 12.3 && 14.2 / 8.0 & 13.9 / 8.0\\
        \midrule
        \checkmark & \checkmark & \checkmark && 5.8 / 12.4 & \textbf{5.7 / 12.1} && 14.3 / 8.0 & \textbf{13.2 / 7.6}\\
        \bottomrule
    \end{tabular}
    \label{tab:swbd_combine}
\end{table*}
\subsection{N-best and Lattice Rescoring}
After pruning the lattices generated by frame-synchronous systems by limiting the beam width and the maximum lattice density, 20-best, 100-best and 500-best hypotheses are obtained from these lattices. The $N$-best hypotheses are then forwarded through the RNNLM, label-sync-LSTM and label-sync-TFM. Each hypothesis has five scores, i.e. AM and LM scores, RNNLM score, and scores from two label-sync systems. By following \eqn{score}, the five interpolation coefficients are found by using the covariance matrix adaptation evolution strategy (CMA-ES)~\cite{Hansen2001CompletelyDS} as a black-box optimisation algorithm to minimise the WER on the dev set. By applying the optimal combination coefficients on the test set, the hypotheses with the highest score are picked.

For AMI-IHM, by comparing columns 4, 5 and 6 in \tbl{ami_combine} along each row, the WER consistently decreases as the $N$-best list size increases. However, the WER improvement from 100 to 500-best is smaller than from 20 to 100-best, i.e. the gain from increasing $N$ reduces for larger $N$. For lattice rescoring (columns 7 and 8 in \tbl{ami_combine}), the improvement from using a higher-order $n$-gram approximation is marginal whereas the lattice density nearly doubles from 4-gram to 5-gram. For both $N$-best and lattice rescoring, \tbl{ami_combine} shows that the WER is lower when combining scores from more models. Furthermore, if just an additional model is to be used for combination with a frame-synchronous system as in the first block of \tbl{ami_combine}, using a label-synchronous model seems to be more effective than an RNNLM, because a label-synchronous model can be viewed as an audio-grounded language model. The WER of the combined system is 25-29\% relative lower compared to a single system in \tbl{single_wer}. With similar lattice densities, lattice rescoring with a 5-gram approximation has a 2\% relative WER reduction over 500-best rescoring.

For SWB-300, 500-best rescoring and lattice rescoring with a 5-gram approximation are reported for Hub5'00 and RT03 sets in \tbl{swbd_combine}. For RT03, the final combined system using lattice rescoring reduces WER by 19-33\% relative compared to single systems in \tbl{single_wer}. Although the 500-best has greater lattice density than the expanded 5-gram lattice, lattice rescoring has 5-8\% relative WER reduction over 500-best rescoring. This is not unexpected because SWB-300 has longer utterances than AMI-IHM on average as shown in \tbl{data}.

\subsection{Analysis}
\begin{figure}[t]
    \centering
    \includegraphics[width=0.9\linewidth]{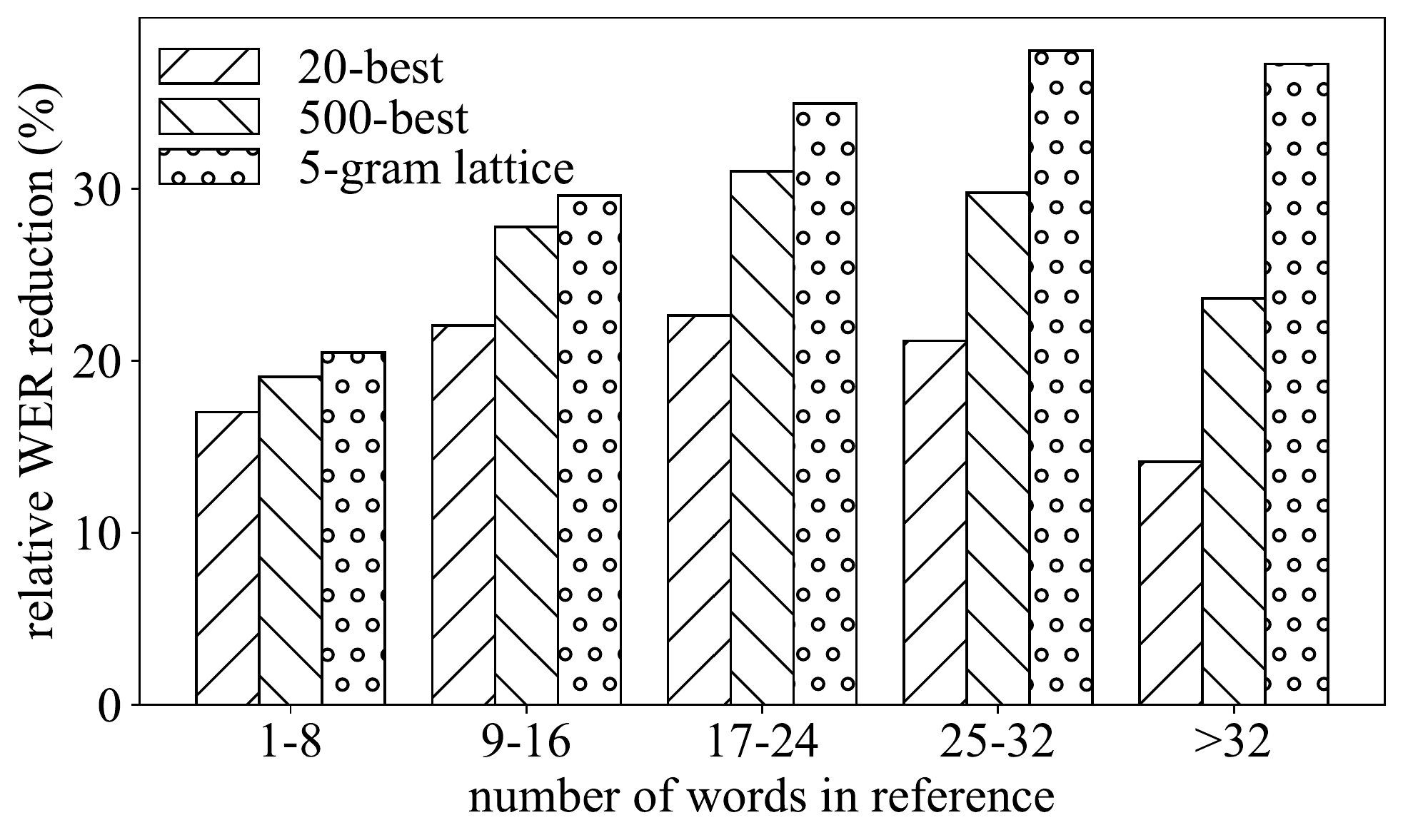}
    \vspace{-1em}
    \caption{Relative WER reduction by utterance length on RT03 for various rescoring methods.}
    \label{fig:werr}
\end{figure}
As discussed in \sect{combine}, lattices are a more compact representation of the hypothesis space that scales well with the utterance length. In \fig{werr}, the relative WER reductions (WERRs) by using 20-best rescoring, 500-best rescoring and lattice rescoring with a 5-gram approximation are compared for different utterance lengths measured by the number of words in the reference. As expected, the gap between $N$-best rescoring and lattice rescoring widens as the utterance length increases. The number of alternatives per word represented by $N$-best is smaller for longer utterances, which explains the downward trend in \fig{werr} for 20-best and 500-best. However, the number of alternatives per word for lattices is constant for a given lattice density. Therefore, the WERR from lattice rescoring does not drop even for very long utterances.

\sect{lsync} compared the complexities of using RNN and Transformer decoders for label-synchronous models. Based on our implementation, speed disparities between the two types of decoders are significant. For 500-best rescoring, label-sync-TFM is about four times faster than label-sync-LSTM. However, label-sync-LSTM is nearly twice as fast as label-sync-TFM for lattice rescoring with a 5-gram approximation. Explicit comparisons between $N$-best and lattice rescoring, and between RNNLM and label-synchronous system rescoring are not made here as they depend on other factors including implementation, hardware, the degree of parallel computation and the extent of optimisation. For example, representing the $N$-best list in the form of a prefix tree~\cite{Sainath2019TwoPassES} or using noise contrastive estimation~\cite{Chen2016EfficientTA} will significantly accelerate $N$-best rescoring using RNNLM.

Under the constraint that no additional acoustic or text data is used, our combined system outperforms various recent results on AMI-IHM~\cite{Kanda2018LatticefreeSM,Sun2021TransformerLM} and SWB-300~\cite{Park2019SpecAugmentAS,Irie2019TrainingLM,Kitza2019CumulativeAF,Wang2020AnIO,Tske2020SingleHA,Saon2021AdvancingRT,Sun2021TransformerLM}. Further improvements are expected if cross-utterance language models or cross-utterance label-synchronous models are used for rescoring~\cite{Irie2019TrainingLM,Tske2020SingleHA,Sun2021TransformerLM}, or combining more and stronger individual systems~\cite{Tuske2021OnTL}.

\section{Conclusion}
\label{sec:conclusion}
In this paper, we have proposed to combine frame-synchronous with label-synchronous systems in a two-pass manner. Frame-synchronous systems are used as the first pass, which can process streaming data and integrate structured knowledge such as lexicon. Label-synchronous systems are viewed as audio-grounded language models to rescore hypotheses from the first pass. Since the two highly complementary systems are integrated at the word level, each one can be trained independently for optimal performance. An improved Lattice rescoring algorithm is proposed for label-synchronous systems, which generally outperform $N$-best rescoring. Label-synchronous systems with RNN decoders are better suited for lattice rescoring while Transformer decoders are more time-efficient for $N$-best rescoring. On AMI and Switchboard datasets, the WERs of the combined systems are around 30\% relatively lower than individual systems.


%





\ifCLASSOPTIONcaptionsoff
  \newpage
\fi



\bibliographystyle{IEEEtran}
\bibliography{refs}
\end{document}